**Dependence of the GRB Lag-Luminosity Relation on Redshift in the Source Frame**

*Walid J. Azzam*

Department of Physics, College of Science, University of Bahrain, Kingdom of Bahrain

wjazzam@sci.uob.bh   or   wjazzam@gmail.com   (FAX  +973-17-449148)

**Abstract**

The lag-luminosity relation for gamma-ray bursts (GRBs) is an anti-correlation between the time lag, $\tau_{lag}$, which represents the delay between the arrival of hard and soft photons, and the isotropic peak luminosity, $L$. In this paper, we use a sample of 43 *Swift* bursts to investigate whether this relation depends on redshift. Both the *z*-correction and the *k*-correction are taken into account. Our analysis consists of binning the data in redshift, $z$, then applying a fit of the form: $\log(L) = A + B\log(\tau_{lag0}/\langle\tau_{lag0}\rangle)$ for each bin, where $\tau_{lag0}$ is the time-lag in the burst's source frame, and $\langle\tau_{lag0}\rangle$ is the corresponding mean value for the entire sample. The goal is to see whether the two fitting parameters, $A$ and $B$, evolve in a systematic way with $z$. Our results indicate that both the normalization, $A$, and the slope, $B$, seem to vary in a systematic way with redshift. We note that although good best-fits were obtained, with reasonable values for both the linear regression coefficient, $r$, and the reduced chi-squared, the data showed large scatter. Also, the number of GRBs in the sample studied is not large, and thus our conclusions are only tentative at this point. A flat universe with $\Omega_M = 0.27$, $\Omega_\Lambda = 0.73$, and a Hubble constant, $H_0 = 70$ km s$^{-1}$ Mpc$^{-1}$ is assumed.

# 1. Introduction

Several gamma-ray burst (GRB) luminosity indicators currently exist. Some are obtained from the light curves, like the lag-luminosity and variability relations [1, 2], while others are obtained from the spectra and include the Amati relation [3–6], the Ghirlanda relation [7], the Yonetoku relation [8, 9], and the Liang-Zhang relation [10]. The importance of these relations lies in their potential use as cosmological probes that might aid in constraining cosmological models [11–13], and as tools that might help in probing the physics of GRBs.

Some investigations, on the other hand, have looked at some inherent problems these relations might suffer from, like the circularity problem and selection effects [14–18]. Other studies have extended the investigation to look at the validity of some of these relations, like the lag-luminosity relation, in the X-ray band and for X-ray flares [19, 20]. However, less attention has been given to the possible redshift evolution of these relations; that is, to the possible dependence of the calibration parameters that appear in these relations on redshift, $z$, as evidenced by the few studies dedicated specifically to this issue [21, 22]. But since these relations are typically calibrated over a wide range in redshift (roughly $0.1 < z < 8$), it becomes important to study their possible dependence on $z$, if they are to prove of any worth as cosmological probes.

The objective of this paper is to look specifically at the possible redshift evolution of one of these luminosity relations – namely, the lag-luminosity relation. The spectral time-lag, $\tau_{lag}$, is defined as the time delay between the arrival of hard and soft photons. Several methods have been used to extract the time-lag, like: the cross-correlation function method, the pulse peak-fit method, and the Fourier analysis method [23]. It was first noted by [1] that there is an anti-correlation between $\tau_{lag}$ and the isotropic peak luminosity, *L*. This correlation was later confirmed by other studies [24–27] and, in fact, used as a cosmological tool [26, 28]. Several studies have also tried to explain the physical origin of the lag-luminosity relation by attributing it to variations in the velocity along the line-of-sight to the GRB [29], or to changes in the off-axis angle [30], or perhaps to a fast radiation cooling effect [31].

In this paper we study the redshift evolution of the lag-luminosity relation by making use of a recent data sample consisting of 43 *Swift* GRBs. The analysis and results are presented in Section 2, which is followed by a discussion in Section 3, and our conclusions are given in the final section.

## 2. Method and Results

In order to properly investigate the possible redshift evolution of the lag-luminosity relation, two corrections have to be kept in mind. The first is the *z*-correction which arises because of time dilation. This is easily accounted for by working in the burst's source frame and using the time-lag in the source frame, $\tau_{lag0} = \tau_{lag}/(1+z)$, instead of the time-lag in the observer's frame, $\tau_{lag}$. The second correction, known as the *k*-correction, is more involved. It has to do with the fact that for bursts with different redshifts, the energy bands in the observer's frame do not map into the same energy bands in the source frame [25]. Correcting for this effect is not an easy task and was recently tackled by Ukwatta et al. [32, 33], who started off by fixing two suitable energy bands in the source frame and then mapping these bands to the observer's frame by using $E_{observer} = E_{source}/(1+z)$. After carrying out this mapping, they used a modified cross-correlation method to extract the appropriate time lags.

The data used in this study is taken from Ukwatta et al. [33] and consists of 43 GRBs. The fixed energy bands in the source frame that they used were: 100-150 keV and 200-250 keV, and they were able to obtain good fits for the lag-luminosity relation. However, no attempt was made to investigate whether the fit parameters, themselves, depend on $z$, This is understandable, since the focus of their paper was not the redshift evolution of the lag-luminosity relation.

Although the entire data sample consists of 43 GRBs, 19 bursts had either negative time-lags (i.e., the photons arrived in the "soft" channel before the "hard" channel), or a "significance" less than $1\sigma$ (see [33] for details), and so were excluded from both the analysis by Ukwatta et al. [33] and our analysis. It should be noted that negative lags are not necessarily unphysical [34], and like [33], we are aware that by omitting them we might be introducing some bias. However, this should not affect our overall conclusions, since what concerns us in this study is whether the fitting parameters vary with $z$, and not their precise values.

Our method consists of binning the data by redshift, then writing the lag relation as:

$$\log(L) = A + B \log(\tau_{lag0}/\langle\tau_{lag0}\rangle) \qquad (1)$$

and extracting the fit parameters $A$ and $B$ for each bin; the goal is to see whether $A$ and $B$ vary in any systematic way with $z$. Note that we normalized $\tau_{lag0}$ to the corresponding mean value for the entire sample, $\langle\tau_{lag0}\rangle = 0.15$ s. This was done in order to avoid introducing any spurious correlations between the two fit parameters.

The isotropic peak luminosity, $L$ (in erg/s), is calculated from the peak flux, $P$ (in erg s$^{-1}$ cm$^{-2}$) using:

$$L = 4\pi d_L^2 P, \tag{2}$$

where $d_L$ is the luminosity distance (in cm), which is obtained assuming a flat universe with $\Omega_M = 0.27$, $\Omega_\Lambda = 0.73$, and a Hubble constant, $H_0 = 70$ km s$^{-1}$ Mpc$^{-1}$. It is perhaps worth clarifying that $L$ refers to the isotropic peak luminosity and not to the bolometric luminosity, and that $P$ corresponds to the observed peak flux for the source-frame energy range 1 keV to 10,000 keV.

The binning was done by fixing the number of bursts per redshift bin. Three bins were used and the number of bursts per bin was 8. **Table 1** shows our results when an unweighted least-squares fit was used. The first three columns show, respectively, the bin number, the number of bursts in the bin, and the redshift range for that particular bin. Columns 4 and 5 show the best-fit values for $A$ and $B$, respectively, along with their 1$\sigma$ errors. The values for the linear regression coefficient, $r$, and the reduced chi-squared values, $\chi_\nu^2$, were also calculated and are shown in columns 6 and 7, respectively. **Table 2** is similar to **Table 1** but shows our results for a weighted least-squares fit in which the errors in both $L$ and $\tau_{lag}$ were taken into account. Both tables show that the goodness of the fits varied from bin to bin, with some bins having very good fits while others had satisfactory fits, which is probably due to both the paucity of points and to the well-known scatter in the lag-luminosity relation [23]. A quick comparison between the two tables shows that both display the same trend for the way $A$ and $B$ vary with redshift.

**Table 1:** *The best-fit values for the normalization, A, and the slope, B, along with their 1σ errors, obtained for different redshift bins, when an unweighted least-squares fit was used. The linear regression coefficient, r, and the reduced chi-squared are also shown.*

| Bin number | No. of GRBs | Redshift range | A | B | r | $\chi_\nu^2$ |
|---|---|---|---|---|---|---|
| 1 | 8 | 0.540–1.091 | 51.94 ± 0.11 | −0.92 ± 0.19 | −0.89 | 0.24 |
| 2 | 8 | 1.101–1.727 | 52.12 ± 0.08 | −0.82 ± 0.12 | −0.94 | 0.13 |
| 3 | 8 | 1.949–3.913 | 52.90 ± 0.12 | −0.04 ± 0.22 | −0.06 | 1.16 |

**Table 2:** *The best-fit values for the normalization, A, and the slope, B, along with their 1σ errors, for the same redshift bins shown in Table 1, but using a weighted least-squares fit. The linear regression coefficient, r, and the reduced chi-squared are also shown.*

| Bin number | No. of GRBs | Redshift range | A | B | r | $\chi_\nu^2$ |
|---|---|---|---|---|---|---|
| 1 | 8 | 0.540–1.091 | 52.06 ± 0.06 | −0.85 ± 0.09 | −0.92 | 0.30 |
| 2 | 8 | 1.101–1.727 | 52.17 ± 0.07 | −0.72 ± 0.10 | −0.95 | 0.15 |
| 3 | 8 | 1.949–3.913 | 53.01 ± 0.13 | −0.07 ± 0.25 | −0.16 | 1.80 |

**Figure 1** shows the best-fit lines for the lag-luminosity relation when the unweighted fit was used, and **Figure 2** shows how the corresponding values for *A* and *B* vary with redshift. The vertical bars in both figures refer to 1σ errors, while the horizontal bars in **Figure 2** show the redshift range of the bin. The exact horizontal location of the points in **Figure 2** was set at the mean redshift value for each bin. **Figure 3** and **Figure 4** are similar to **Figure 1** and **Figure 2**, respectively, but with the use of a weighted least-squares fit (see **Table 2**). Both **Figure 2** and **Figure 4** show a clear increase of the normalization, *A*, and the slope, *B*, with *z* especially at high redshift. Similar results were obtained when 4 bins were used instead of 3. Thus, the lag-luminosity relation seems to vary with redshift through a systematic dependence of the fitting parameters on *z*. However, we cannot make strong or conclusive statements at this point since the number of bursts in the sample studied is small, but if future studies with larger data samples do confirm our results, then one has to be cautious in using the lag-luminosity relation with bursts that span a wide range in redshift.

## 3. Discussion

In this section we would like to put our study in proper context by comparing it to what has been done by others. As mentioned earlier, only a few studies have specifically targeted the issue of redshift evolution of GRB luminosity relations. Among these is the paper by [21] in which a sample of 48 GRBs was used to investigate the redshift evolution of the $E_p$-$E_{iso}$ relation (the Amati relation), where $E_p$ is the peak energy obtained from the $E^2N(E)$ versus $\nu f_\nu$ distribution and $E_{iso}$ is the isotropic energy. The author of that paper found evidence that this relation gets steeper with redshift, and concluded that the Amati relation seems to evolve with redshift. A subsequent investigation by [16] extended the study done by [21] by enlarging the data sample to 76 bursts. Although they confirmed the results of [21] for the 48 bursts, when all 76 bursts were used the redshift evolution disappeared, and hence their conclusion was that what [21] had found was probably due to low statistics.

The paper by [22] investigated the possible redshift evolution of the lag-luminosity relation and is thus more relevant to our current paper, however, both their approach and data sample are different from ours. Using the Yonetoku relation, they extracted the redshifts, $z_Y$, of 565 BATSE bursts and compared them to the redshifts, $z_{lag}$, extracted using the lag-luminosity relation. To their surprise, the two sets of redshifts did not correlate well, and in order to bring them into agreement the lag-luminosity relation itself had to evolve with redshift.

Although the current paper confirms what was found by [22], one should keep in mind that the data sample used is not large and the lag-luminosity relation shows considerable scatter [23], hence, like [21] we might be governed by low statistics. However, if future investigations do confirm our results, then one should be cautious in using the lag-luminosity relation as a cosmological probe, especially if a wide range in redshift is involved.

## 4. Conclusions

A sample of 43 *Swift* GRBs was used to investigate the possible redshift evolution of the lag-luminosity relation. Our analysis indicates that the normalization, *A*, and the slope, *B*, vary with *z* especially at high redshift. Thus, the lag-luminosity relation does seem to evolve with redshift, however, our conclusions are only tentative at this point, since our study is limited by low statistics and noticeable scatter in the lag relation.

**Acknowledgements**

The author would like to thank H. A. Eid for useful discussions and input concerning this paper.

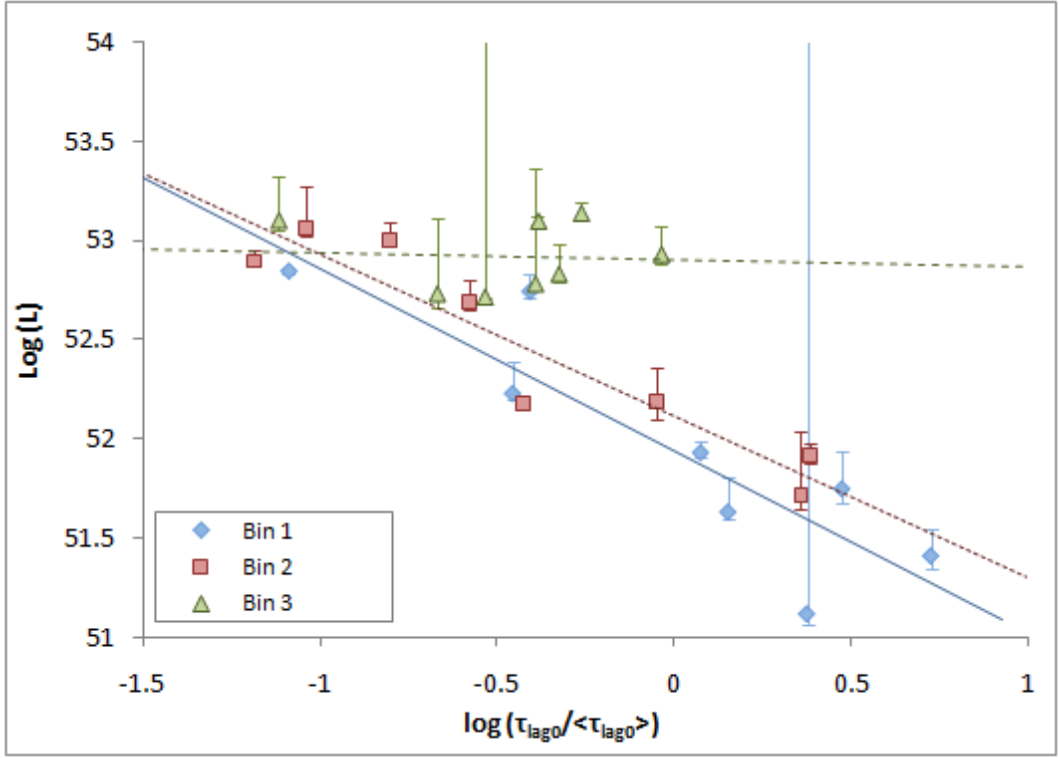

**Figure 1** The isotropic peak luminosity, *L*, plotted versus the normalized time-lag in the burst's source frame. The best-fit lines for an unweighted least-squares fit are shown for the three redshift bins used, where bin 1 represents the lowest redshift range. The vertical bars refer to 1$\sigma$ errors.

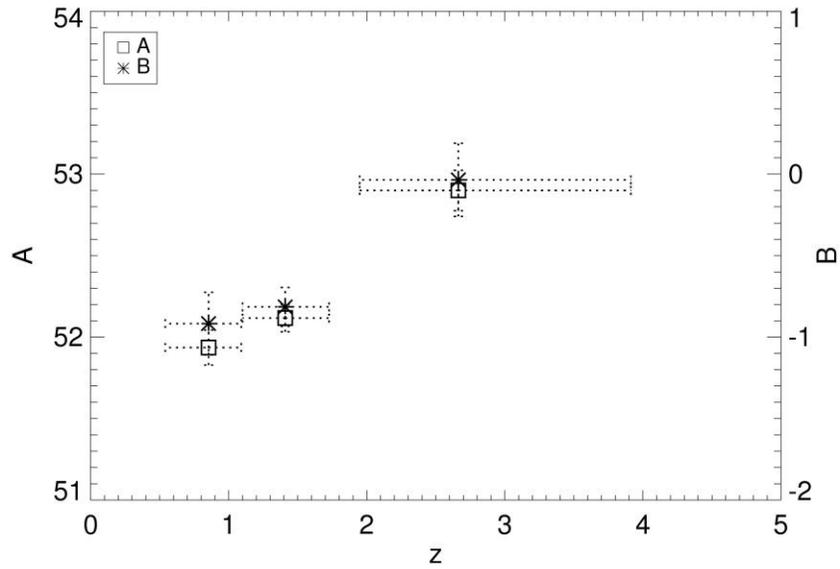

**Figure 2** The normalization, *A*, and the slope, *B*, plotted versus the redshift for the unweighted fits shown in **Figure 1**. The vertical bars refer to 1σ errors, while the horizontal bars show the redshift range of the bin.

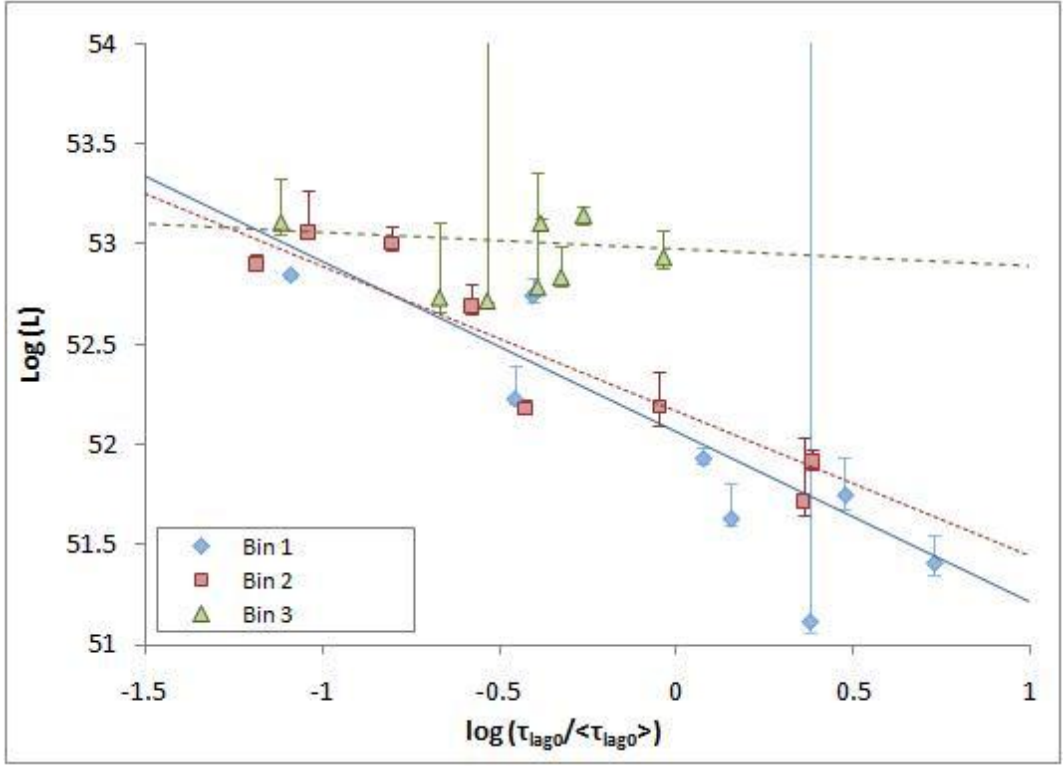

**Figure 3** The isotropic peak luminosity, *L*, plotted versus the normalized time-lag in the burst's source frame. The best-fit lines for a weighted least-squares fit are shown for the three redshift bins used, where bin 1 represents the lowest redshift range. The vertical bars refer to 1σ errors.

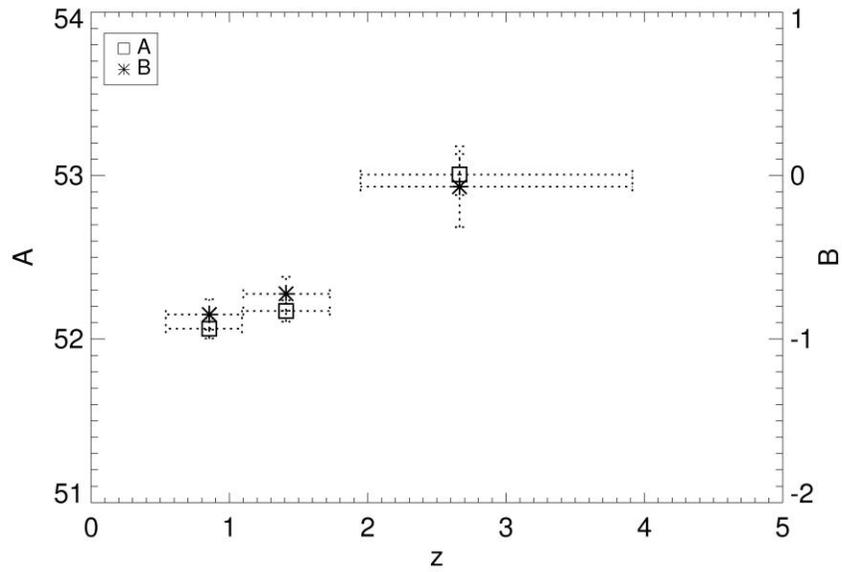

**Figure 4** The normalization, *A*, and the slope, *B*, plotted versus the redshift for the weighted fits shown in **Figure 3**. The vertical bars refer to 1σ errors, while the horizontal bars show the redshift range of the bin.

15